# ON THE GEOMETRIC, ENERGETIC AND MAGNETIC PROPERTIES OF THE MONOMER, DIMER AND TRIMER OF $NIFE_2O_4$.

José Burgos[a], Luis Seijas[c], Pedro Contreras[b], Rafael Almeida[c,*]

[a]*Departamento de Física, Facultad de Ciencias, Universidad de Los Andes, Mérida, Venezuela.*

[b]*Centro de Física Fundamental, Departamento de Física, Facultad de Ciencias, Universidad de Los Andes, Mérida, Venezuela.*

[c]*Grupo de Procesos Dinámicos, Departamento de Química, Facultad de Ciencias, Universidad de Los Andes, Mérida, Venezuela.*

[*]Corresponding author: Rafael Almeida, Grupo de Procesos Dinámicos, Departamento de Química, Facultad de Ciencias, Universidad de Los Andes, Mérida, Venezuela. E-mail: mata@ula.ve

**Abstract.** In this work, by employing Density Functional Theory, we compute and discuss some geometric, energetic and magnetic properties of the monomer, dimer and trimer of $NiFe_2O_4$. The calculations are performed at the UDFT/ B3LYP level of calculation, by employing the LANL2DZ effective pseudo potential. The results of the Mulliken spin densities and the spin polarization will be presented. Finally the outcome of the system density of states is considered.
.

## 1. Introduction

Understanding the magnetic properties of small clusters is important, not only from the fundamental point of view, but also because of their potential applications [1]. However, due to the reduction in the number of nearest neighbors, usually the bulk picture is inadequate for the description of these systems, and in many cases, it is necessary to resource to theories or methodologies developed for molecular system [2].

Thanks to their phenomenological behavior and their industrial applications, Ferrites constitute a group of magnetic systems that have captured a great deal of interest [3]. Among this group of magnetic materials, Nickel ferrites (*$NiFe_2O_4$*) is an insulating magnetic oxides with high magnetic order and a saturation magnetization, that due to its low microwave loss, low magnetic anisotropy and low magnetostriction have been used for RF/microwaves applications [4]. These compounds have inverse spinel structure. The spinel has a cubic arrangement with two non-equivalent cations *A, B,* and general formula

$AB_2O_4$. The spinel structure contains two nonequivalent cation sites: i) tetrahedral *A* sites, and ii) octahedral *B* sites. In normal spinel structure, all the *A* sites are occupied by the divalent cation, while all *B* sites are occupied by the trivalent cation. On the other hand, in the inverse spinel structure, trivalent cations occupy all the tetraedric *A* sites, and 50 % of the octaedric *B* sites, while the remaining 50% of the *B* sites are occupied by the divalent cations: $Fe_A[NiFe]_BO_4$ [5]. The compounds $NiFe_2O_4$ crystallized in the inverse spinel structure with space group Fd3m. From the magnetic point of view, there is ferromagnetic *A-A* interaction, and a antiferromagnetic super exchange interaction, via oxygen, between *A* and *B* nearest neighbor sites, resulting in the dominant antiparallel alignment of these sites [6]. These magnetic states involve the existence of quasilocal magnetic moments, that remain mutually aligned due to the strong interatomic exchange interactions, $J_{AB}$. Thus, structural modifications would change the magnetic moment relative orientations, and in this manner, the interactions among these dipoles; and therefore, affecting the system magnetic behavior. In the bulk, the strong intra-site Coulomb repulsion between d-electrons splits the partially filled d-band into upper and lower Hubbard bands, resulting simultaneously in an insulating electronic ground state and a local magnetic moment: Mott insulator [7].

Many of the initial theoretical calculations on the spinel ferrites electronic structure have focused on the magnetite ($Fe_3O_4$). [8] This material may be seen as the parent compound of the spinel-type ferrites. Thus, a compound, as $NiFe_2O_4$, is obtained by substituting the magnetite $Fe^{2+}$ cation by other transition metal *3d* divalent cation, i.e, $Ni^{2+}$. Previous works on the NiFe2O4 ferrite have been developed at the DFT level under the local spin density approximation (LSDA) [9], the linear muffin-tin method [10], hybrid functionals [11] and the generalized gradient approximation (GGA) together with the Hubbard+U correction [12]; however, as far as we know, no studies have been carried out focusing on the geometric, energetic and magnetic properties of $NiFe_2O_4$ small clusters. Thus, the goal of this work is to contribute in this direction, by studying these properties for the $NiFe_2O_4$ monomer, dimer and trimer.

In the next section, a summary of the computational methodology is provided, followed by the geometrical and energetic results. Since, as mentioned before, the interaction between the spins in ferrites compounds occurs through a superexchange mechanism [6], with the purpose of numerically describing the magnetic interactions, the results of the Mulliken spin densities and the spin polarization will be presented. Finally the outcome of the system density of states is considered.

**Computational details**

The $NiFe_2O_4$ monomer, dimer and trimer are studied at the UDFT level of calculation,

employing the Becke's three-parameter hybrid functional (B3) [13] and the Lee-Yang-Parr correlation functional (LYP) [14], using the Gaussian03 program [15]. The base employed in this study is the LANL2DZ effective pseudo potential [16]. The LANL2DZ pseudo potential belongs to the operators that consider an Effective Core Potential (ECP) for the inner electrons. Typically, these potentials are formed by the sum of products of polynomial radial functions, Gaussian radial functions and angular momentum projection operators [15].

## 2. Results and discussion

### 2.1 Geometrical and Energetic Results

Geometrical optimization was performed to find the ground state structures, the result of which will be used to discuss the magnetic properties. The outcomes for the $NiFe_2O_4$ monomers, dimers and trimers geometry optimization are displayed in Fig. 1. Each of these structures belongs to the *Cs* symmetry group. Notice that the Ni atoms are tetraedrically bonded to oxygen ones; however, for the small clusters considered here, it the Fe coordination pattern is not clear yet. For each case, all the multiplicities were considered and their energies were calculated. The results for the most stable cases, together with their multiplicities and the energy difference between the ground state and the first excited one are shown in Table 1.

FIG. 1. Optimized geometries of the $NiFe_2O_4$ monomer, dimer and trimer.

a) $(NiFe_2O_4)$

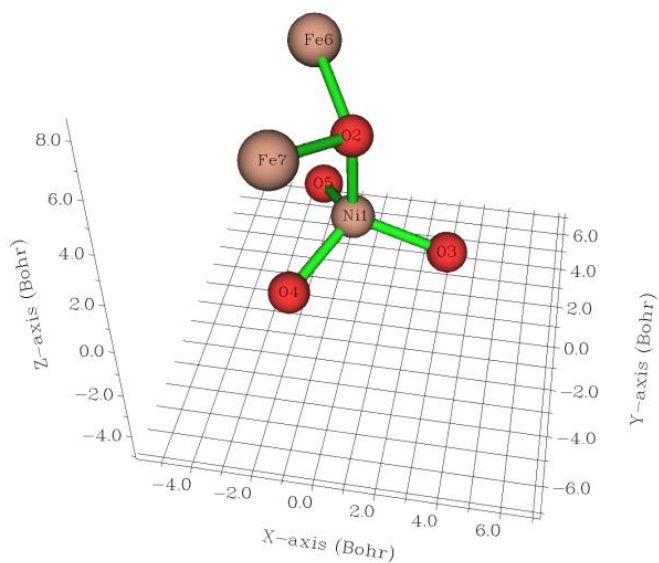

b) (NiFe$_2$O$_4$)$_2$

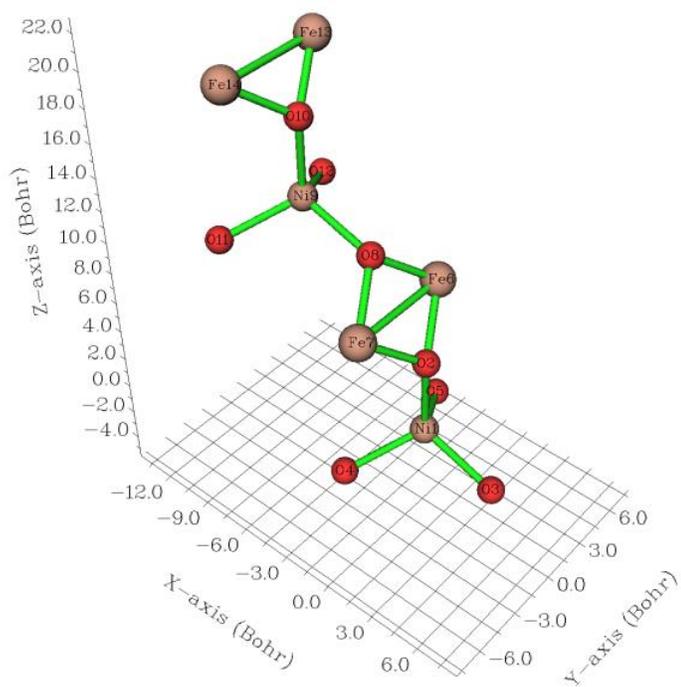

a) (NiFe$_2$O$_4$)$_3$

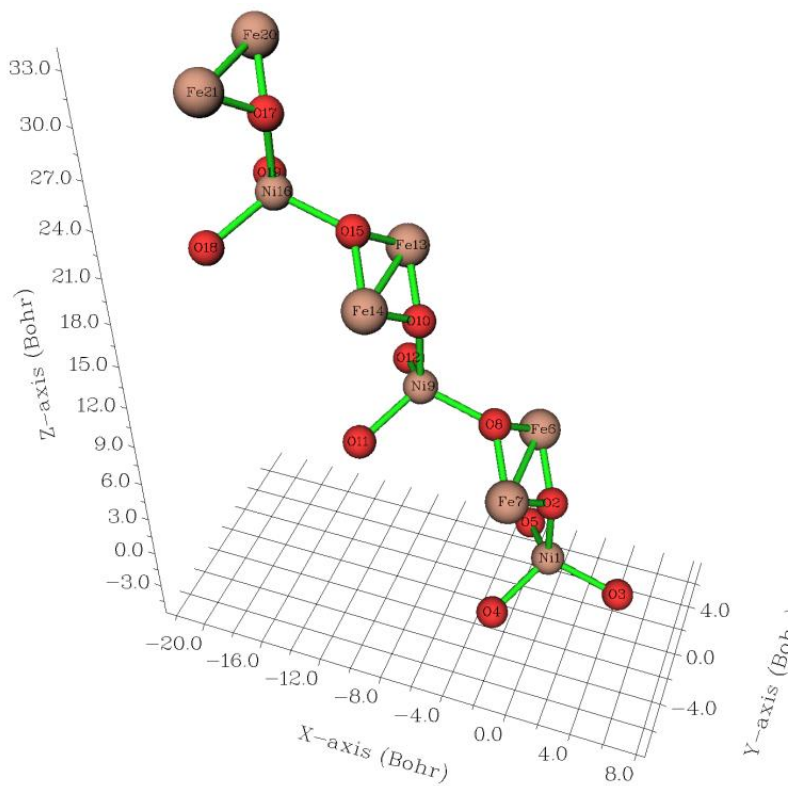

TABLE 1

Ground state energy and multiplicity for the NiFe2O4, (NiFe2O4)₂ and (NiFe2O4)₃. The fourth column corresponds to the energy difference between the ground and first excited states.

| Structure | Multiplicity | Energy (kcal/mol) | $\Delta$E(kcal/mol) |
|---|---|---|---|
| (NiFe2O4)₁ | 13 | -449822.82 | 2,95 |
| (NiFe2O4)₂ | 25 | -899454.64 | 0,75 |
| (NiFe2O4)₃ | 37 | -1349230.15 | 1,00 |

As should be expected, it was found that larger the multiplicity, most stable is the system. For the monomer case, the most stable state energy is -449822.82 kcal/mol and its multiplicity is 13, the first excited state has multiplicity 3, and the energy difference, $\Delta$E, respect to the ground state is approximately 3 kcal/mol. For the dimer, the ground state has multiplicity 25, and its energy is -899454.64 kcal/mol. For this case, the first excited state, with multiplicity 17, has energy barely about 1 kcal/mol larger than that of the ground state. The trimer most stable state has energy of -1349230.15, and again, the energy difference with the state with multiplicity 23 is just 1 kcal/mol. Let us emphasize that the energy differences between the two more stable state is quite small, and become smaller when the system size increases.

## 2.2 Spin densities and magnetic properties

For each of the monomer, dimer and trimer multiplicities, Table 2 shows the computed value of the spin quantum number S, the $S^2$ operator eigenvalues (in $\hbar$ units), the number of electrons with spins $\alpha$ and $\beta$, and the spin polarization fraction. The S(theory) and $S^2$(theory) correspond to the expected values in each case. The difference between the computed S and $S^2$ values and those of S (theory) and $S^2$ (theory), respectively, provides a measure of the spin contamination in each case. As has been previously reported [??], the results show that smaller the multiplicity larger the extension of the spin contamination, pointing out that, for the ground and first excited states, the calculated spin quantum numbers S and the $S^2$ operator eigenvalues are quite close to their expected values, indicating the degree of reliability of our results.

This Table also displays the outcome for the spin polarization fraction $\xi$, which is defined as the relative difference between the $\alpha$ and $\beta$ electrons [17],

$$\xi = \frac{\#\text{electrons } \alpha - \#\text{electrons} \beta}{\#\text{electrons} \alpha + \#\text{electrons} \beta}$$

For all the systems is found that ξ decreases with the multiplicity, becoming negligible for the smallest multiplicity cases.

## TABLE 2

Results for the calculated values of the spin quantum numbers S, the $S^2$ eigenvalues (in $\hbar$ units), the number of spin α and β electrons, and the spin polarization fractions. The S(theory) and $S^2$ (theory) correspond to the expected values in each case.

a) $(NiFe2O4)_1$

| Multiplicity | S | S(theory) | $S^2$ | $S^2$ (theory) | #electrons α | #electrons β | ξ (spin polarization Fraction) |
|---|---|---|---|---|---|---|---|
| 13 | 6.05 | 6 | 42.65 | 42 | 47 | 35 | 0.15 |
| 9 | 4.10 | 4 | 22.13 | 20 | 45 | 37 | 0.09 |
| 3 | 2.18 | 1 | 8.03 | 2 | 42 | 40 | 0.02 |

b) $(NiFe2O4)_2$

| Multiplicity | S | S(theory) | $S^2$ | $S^2$ (theory) | #electrons α | #electrons β | ξ (spin polarization Fraction) |
|---|---|---|---|---|---|---|---|
| 25 | 12.03 | 12 | 156.80 | 156 | 94 | 70 | 0.15 |
| 17 | 8.29 | 8 | 77.02 | 72 | 90 | 74 | 0.09 |
| 15 | 7.37 | 7 | 62.21 | 56 | 89 | 75 | 0.08 |
| 11 | 5.68 | 5 | 37.94 | 30 | 87 | 77 | 0.06 |
| 7 | 4.22 | 3 | 21.85 | 12 | 85 | 79 | 0.03 |
| 5 | 3.64 | 2 | 16.87 | 6 | 84 | 80 | 0.02 |
| 1 | 0.00 | 0 | 0.00 | 0 | 82 | 82 | 0.00 |

c) $(NiFe2O4)_3$

| Multiplicity | S | S(theory) | $S^2$ | $S^2$ (theory) | #electrons α | #electrons β | ξ (spin polarization fraction |
|---|---|---|---|---|---|---|---|
| 37 | 18.02 | 18 | 342.96 | 342 | 141 | 105 | 0.15 |
| 33 | 16.09 | 16 | 274.92 | 272 | 139 | 107 | 0.13 |
| 29 | 14.18 | 14 | 215.82 | 210 | 137 | 109 | 0.11 |
| 27 | 13.21 | 13 | 188.54 | 182 | 136 | 110 | 0.10 |
| 23 | 11.34 | 11 | 140.13 | 132 | 134 | 112 | 0.08 |
| 19 | 9.49 | 9 | 99.59 | 90 | 132 | 114 | 0.07 |
| 17 | 8.61 | 8 | 82.83 | 72 | 131 | 115 | 0.06 |
| 13 | 6.98 | 6 | 55.76 | 42 | 129 | 117 | 0.04 |
| 9 | 5.39 | 4 | 34.50 | 20 | 127 | 119 | 0.03 |
| 3 | 4.00 | 1 | 20.01 | 2 | 124 | 122 | 0.01 |

Table 3 displays the Mulliken spin density results for the largest multiplicity cases. The atom numbers correspond to those assigned in Figure 1. From that figure, two kinds of oxygen atoms could be distinguished, the first one, where the oxygen acts as a bridge between the Ni and Fe atoms (O2 in TABLE 3-a, O2 and O8 in b, and O2, O8 and O15 in c), and the second kind, in which the oxygens are only bonded to the Ni atoms (O3, O4 and O5 in TABLE 3-a, O3, O4, O5, O11 and O12 in b, and O3, O4, O5, O11, O12, O18 and O19 in c). The results show that the second group oxygens have larger spin densities than those in the first one, and in both cases, the density increases from monomer to dimer, remaining approximately constant for the trimer. From this Table is also observed that the largest spin densities are found for the Fe atoms, with their values increasing with the number of neighbors. Thus, for the trimer, the Fe13, Fe14 densities are larger than those of the Fe6, Fe7, which are larger than the F20, Fe21 spin densities. As mentioned before, the Ni atoms are bonded tetraedrically bonded to the oxygens; however, two groups of Ni atoms are found, the first is formed by the Ni closer to one of the extreme of the molecular system (Ni1), which has the smallest density value (~1.20), while for the second group, each Ni has four Fe atoms as second neighbors, and their density is about 1.30.

TABLE 3

Mulliken atomic spin densities for the ground state systems. The atom numbers correspond to those indicated in Figure 1.

a) $(NiFe2O4)_1$, Multiplicity 13

|   | Atomic | spin densities |
|---|---|---|
| 1 | Ni | - 1.56 |
| 2 | O | 0.28 |
| 3 | O | -1.47 |
| 4 | O | -1.34 |
| 5 | O | -1.37 |
| 6 | Fe | 2.99 |
| 7 | Fe | 2.99 |

b) $(NiFe2O4)_2$, Multiplicity 25

|   | Atomic | spin densities |
|---|---|---|
| 1 | Ni | 1.21 |
| 2 | O | 0.45 |
| 3 | O | 1.62 |
| 4 | O | 1.56 |
| 5 | O | 1.57 |
| 6 | Fe | 3.49 |
| 7 | Fe | 3.49 |
| 8 | O | 0.52 |
| 9 | Ni | 1.30 |
| 10 | O | 0.26 |
| 11 | O | 1.39 |
| 12 | O | 1.38 |
| 13 | Fe | 3.13 |
| 14 | Fe | 3.13 |

c) (NiFe2O4)$_3$, Multiplicity 37

| | Atomic | spin densities |
|---|---|---|
| 1 | Ni | 1.19 |
| 2 | O | 0.46 |
| 3 | O | 1.63 |
| 4 | O | 1.56 |
| 5 | O | 1.60 |
| 6 | Fe | 3.49 |
| 7 | Fe | 3.49 |
| 8 | O | 0.51 |
| 9 | Ni | 1.28 |
| 10 | O | 0.38 |
| 11 | O | 1.39 |
| 12 | O | 1.39 |
| 13 | Fe | 3.55 |
| 14 | Fe | 3.48 |
| 15 | O | 0.50 |
| 16 | Ni | 1.31 |
| 17 | O | 0.26 |
| 18 | O | 1.39 |
| 19 | O | 1.38 |
| 20 | Fe | 3.14 |
| 21 | Fe | 3.14 |

In order to shed some light on these results, in Figure 2, for each of the considered cases, their spin polarization density contours are exhibited. From there is clear that the Fe atoms have a ferromagnetic behavior, indicated by blue contours, while the Ni atoms have an anti-ferromagnetic one, corresponding to pink contours. Thus, the Fe-Ni antiferromagnetic super exchange interaction, occurs via the oxygens that act as a bridge between the Ni and Fe atoms (O2 in TABLE 3-a, O2 and O8 in b, and O2, O8 and O15 in c), resulting this in smaller density concentrations over those oxygen atoms. From that Figure is also clear that the Ni atoms near the extreme only interact with a pair of Fe atoms, whereas the other Ni atoms interact with two of those pairs, yielding, these two kinds of interactions, different spin densities over the Ni atoms involved. From Figure 2, one also can observe that, in addition to the super exchange interaction with the Ni, there is also a direct ferromagnetic interaction between the Fe nearest neighbor atoms, resulting in the dominant spin density localization over these atoms.

As a consequence of the magnetic behavior discussed before, the NiFe2O4 monomer, dimer and trimer show permanent magnetic dipole moments, whose magnitudes are calculated to be ???, $$$, and ***, respectively.

FIG. 2. Spin Polarization densities contours for the NiFe$_2$O$_4$ monomer, dimer and trimer. Blue contours correspond to ferromagnetic states, while pink ones correspond to anti-ferromagnetic states

a) (NiFe$_2$O$_4$)

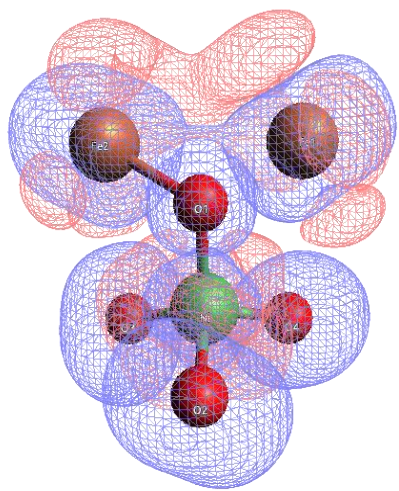

b) (NiFe$_2$O$_4$)$_2$

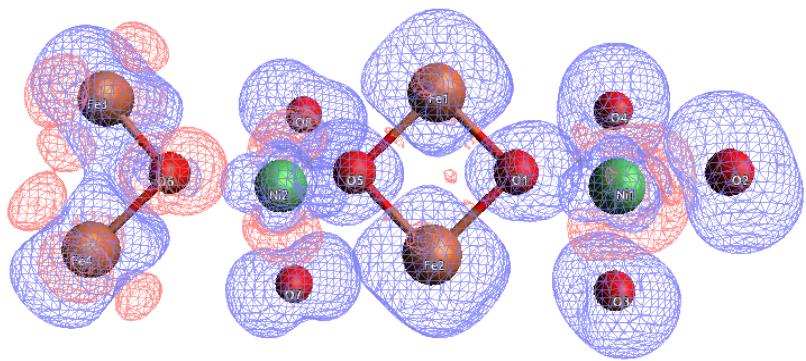

c) (NiFe$_2$O$_4$)$_3$

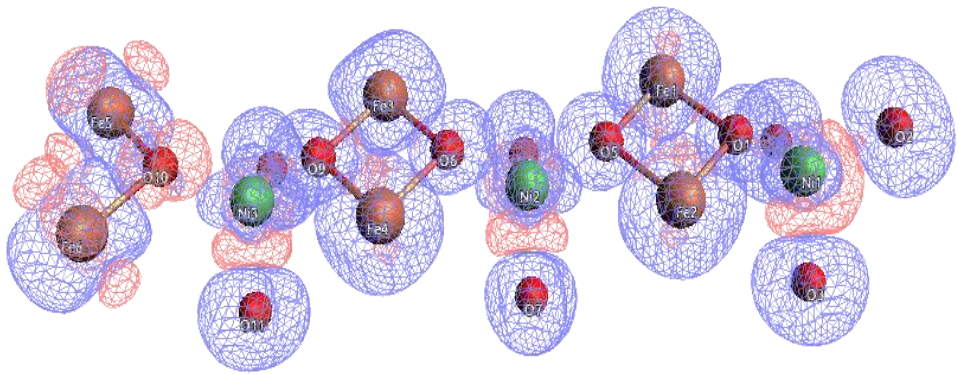

## 2.3 Density of states and electronic structure

The density of states (DOS) for the $NiFe_2O_4$, $(NiFe_2O_4)_2$ and $(NiFe_2O_4)_3$ ground states (spin multiplicity 13, 25 and 37 respectively) is calculated and the results are displayed in Figure 3. There both, the spin alpha DOS, upper part of each figure, and spin beta DOS (lower part) are shown. For all the cases, the total DOS associated with occupied states is larger for the alpha electrons, while the opposite is observed for the unoccupied ones. For the monomer, it is observed that the Ni d-states have lower energies than those of Fe (although for the alpha electrons a small DOS is observed near the HOMO energy), with those states showing small overlap between them. On the other hand, the O p-DOS band extends all over the considered energy range, showing a significant overlap with both Fe and Ni states, which is in agreement with the O function in the superexchange interaction. Also is noticeable that for the unoccupied states the largest DOS corresponds to the beta electrons. For the dimer, the alpha electron Ni d-states and Fe d-states strongly overlap between them, and with the O p-states. However for the beta electrons, it is observed that the Ni d-states have lower energies than those of the Fe, although in this case those states overlap at about -8 eV. For the unoccupied states, the extension of these orbital overlap grows, respect to the occupied ones. Finally for the trimer, the alpha electrons Fe d-states and the O p-states split into two groups, the Fe higher energy one, and the O higher energy group, almost perfectly overlap with the Ni d-states, representing the states that participate in the superexchange interaction. Thus, the Fe d-states splitting is associated with the two kinds of magnetic interactions where the Fe-atoms participate: The direct paramagnetic interaction, and the superexchange antiferromagnetic one. Something similar is found for the beta electrons; nevertheless in this case, for the occupied states the splitting and the overlap are such that the extension is smaller than that observed for the alpha electrons; whereas again, for the unoccupied states the opposite trend is obtained.

Figure 3 also shows the state energies, represented by vertical lines. From those results it is found that the $NiFe_2O_4$ LUMO/HOMO Energy gap is 2.8518 eV for the α-electrons and 1.8317 eV for the β ones. For the $(NiFe_2O_4)_2$ this difference is 2.2503 eV for α-electrons and 0.4828 eV for the β ones, while for the $(NiFe_2O_4)_3$, the gap for the α and β electrons is 1.7986 eV and 0.4663 eV respectively.

FIG. 3. Density of states for the $NiFe_2O_4$ monomer, dimer and trimer. For each case, the top panel represents the alpha electrons DOS results, while the bottom one corresponds to the beta-electrons. In addition to the total DOS, the Ni 3d, Fe 3d and O 2p DOS are also represented.

a) (NiFe$_2$O$_4$)

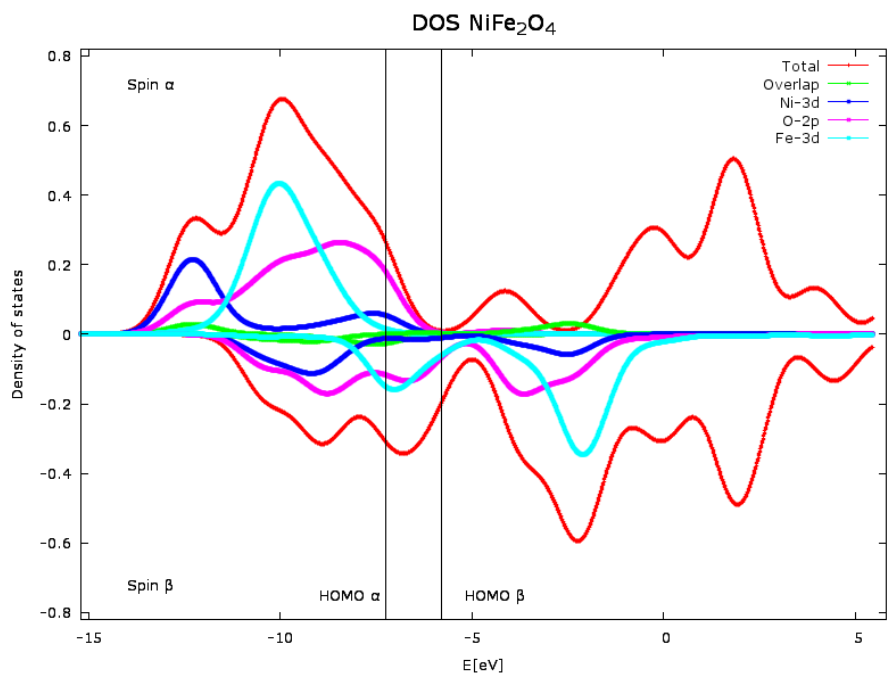

b) (NiFe$_2$O$_4$)$_2$

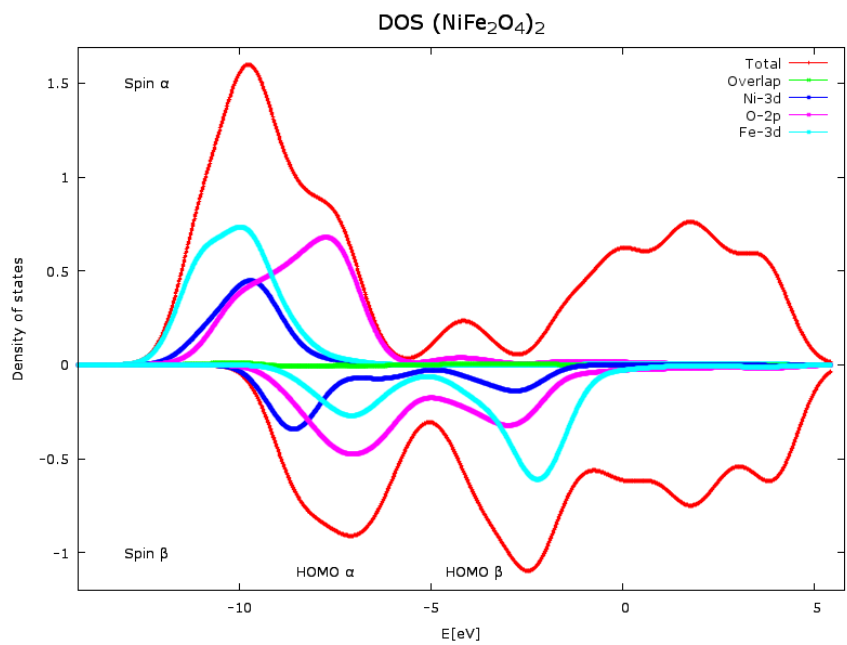

c) (NiFe$_2$O$_4$)$_3$

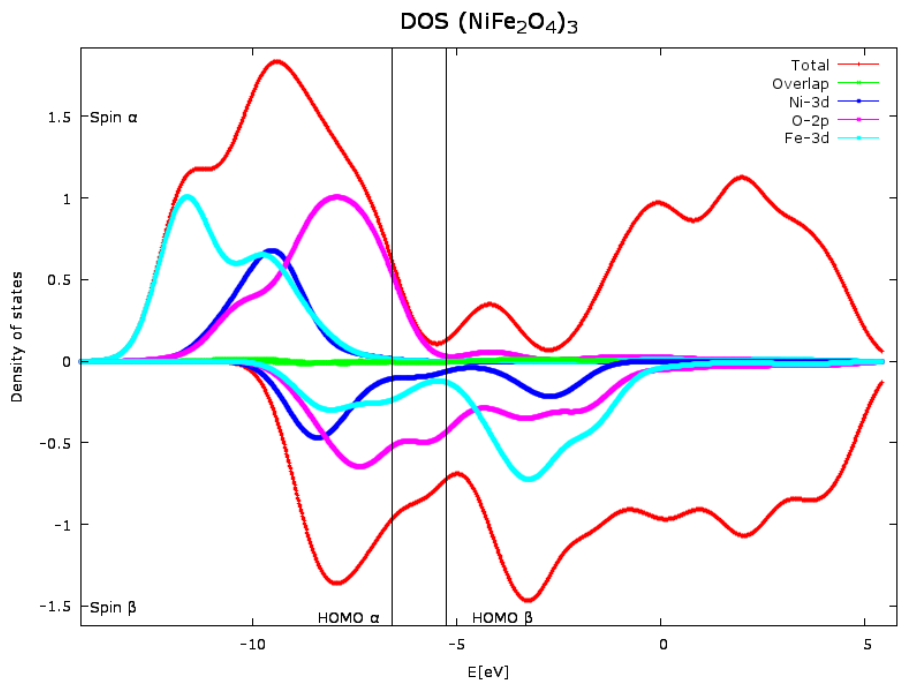

## 3. Summary

Understanding the magnetic properties of small clusters or nanomaterial is an active research area. Among these materials, due to its potential applications, the ferrites have extensively studied. In this direction, in this work, by employing the LANL2DZ effective pseudo potential, we have explored the geometric, energetic and magnetic properties of the $NiFe_2O_4$ monomer, dimer and trimer at the UDFT/B3LYP level of calculation. It was found that each of the structures belongs to the *Cs* symmetry group, while the energy results show that larger the multiplicity, most stable is the system, and that the energy difference, between the first excited state and the ground state is smaller than 3 kcal/mol. Since it was also found that smaller the multiplicity is, larger the extension of the spin contamination; one can conclude that the results for the ground and first excited states are reliable. The results also indicate that for the considered systems, the spin polarization fraction decreases with the multiplicity, becoming negligible for the smallest multiplicity cases. It was found that the DOS structure reflects the kind of magnetic interactions present in the $NiFe_2O_4$ molecular systems, particularly, the direct paramagnetic interaction, and the superexchange antiferromagnetic ones. The outcome of the calculations performed here, confirms the potential of the DFT for studying small magnetic clusters.

**Acknowledgment**

The authors gratefully acknowledge the financial support of the CDCHTA-ULA, through the project number C-1924-15-08-EM. One of us, RA, also acknowledge the financial support from the Spanish *Ministerio de Ciencia e Innovación,* Grant No. FIS2010-21282-C02-02. The computational calculations were partially performed at the Multidisciplinary Center for Scientific Computing and Visualization of the Universidad de Carabobo (CeMViCC).